\documentstyle{article}
\begin{document}
\title{
\bf {Introducing spin\\ to\\ classical phase space}}
\author{
       {\bf J. Le\'on}\thanks{e-mail: leon@laeff.esa.es} \,  and
       {\bf J. M. Mart\'{\i}n}\thanks{e-mail: jmm@laef.esa.es} \\
   Laboratorio de Astrof\'{\i}sica Espacial y F\'{\i}sica Fundamental, INTA\\
        Ap. 50727, E 28080 MADRID, Spain\\  and \\
         Instituto de Matem\'aticas y F\'{\i}sica Fundamental, CSIC\\
        Serrano 123, E 28001 MADRID, Spain}
\date{December 26, 1995}
\maketitle
\vspace{-9.5cm}
 \hfill LAEFF 95/26
\vspace{8cm}
\begin{abstract}
The kinematic degrees of freedom of spinning particles are analyzed and an
explicit construction of the phase space and the simplectic structure that
accomodates them is presented. A Poincare invariant theory of classical
spinning particles that generalizes the work of Proca and Barut to arbitrary
spin is given using spinor variables. Second quantization is naturally
connected to the unphysical nature of zitterbewegung. Position variables can
not be disentangled from spin in a canonical way, nor can the phase space be
reduced to the usual description $(x,p)$ and a vector spin.
\end{abstract}
Pacs:  03.20.+i, 03.65.Sq, 03.30.+p, 11.30.Cp

\newpage

Classical Mechanics
is based on the idea of the point particle endowed  with mass. It
successfully determines the evolution of the particle by  the sole
virtue of the Newton Principles. These have been formulated in a
variety of ways of which in this paper we will use the economical and
elegant Action Principle. Assuming that the action $A$
describes the motion of a free particle through the Euler-Lagrange
equations, we can impose $\delta A=0$ for translations and
Lorentz transformations to ensure that any pair of independent
inertial observers will detect the same motion of the particle. That
this gives the Galileo principle (or its equivalent, the First Newton
Law) can be seen as follows: Be $x^{\mu} , \dot{x}^{\mu}$ and $\tau$
the coordinates, velocities and proper time of  the particle
($\dot{x}^{\mu} ={dx^{\mu} / d\tau}$). The invariance of the  action
implies 
\begin{equation} \delta A=(p_{{\mu}} \delta
x^{{\mu}}){\arrowvert}^{\tau_2}_{\tau_1}   =0, \, p_{{\mu}} \equiv 
\frac{\partial L}{\partial
\dot{x}^{{\mu}}} 
\label{1} 
\end{equation} 
where $\delta x^{\mu}
=\epsilon^{{\mu}}$ or  $\delta x^{\mu} =\epsilon^{\mu}_{\nu} x^{\nu}$
for translations or Lorentz transformations respectively. From the
above one  obtains the conservation of momentum $p_{\mu}$ and angular
momentum  $L_{{\mu}{\nu}} =x_{\mu} p_{\nu} -x_{\nu} p_{\mu}$. In
practice there are only seven  independent  conserved quantities
$p_{\mu}$ and $L_{{\mu}{\nu}} p^{\nu}$, because  
$L^*_{{\mu}{\nu}}p^{\nu}$, where $L^*$ is the dual of $L$, vanish by 
construction. 
The object  $X_{\mu}=\frac{1}{p^2} L_{{\mu}{\nu}} p^{\nu}$ gives the part of 
$x$ orthogonal to $p$, so  the particle trajectory is
simply: 
\begin{equation} x_{\mu} = X_{\mu} + f(\tau) p_{\mu} \label{2} \end{equation} 
where
$f(\tau)$ is a scalar function of the proper time.  From this one gets:
\begin{equation} 
\frac{\vec{x}(\tau) -\vec{x}(0)}{x^0(\tau) -
x^0(0)}=\frac{\vec{p}}{p^0}= \vec{v}, \label{3} \end{equation} 
which is the sought
Galileo Principle written in the most accesible way to an inertial
observer. Summarizing, the trajectory is a straight line  completely
determined by the six constants of motion $p_{{\mu}}/\sqrt{p^2}$  and
$X_{{\mu}}$. The particle travels along the  trajectory with constant
speed $v$. The physical requirement that $\tau$ be the proper time,
i.e. $d\tau^2 =dx_{\mu} dx^{\mu}$, or $\dot{x}^2 =1$,  gives $\dot{f}
=1/m$, where $m=\sqrt{p^2}$ is the particle mass.  In addition, the
observer can set up time -adjusting it to the particle time- by fixing
the initial condition $f(0)$. This derivation is independent of  the
explicit form chosen  or the lagrangian, but a consequence of
combining the action principle with the invariance under Poincare
transformations. 

Classical spinning particles are typically described~\cite{corben} like
points endowed with a mass and a spin angular momentum $S_{\mu \nu}$. 
Along with $p$ the total angular momentum
\begin{equation} J_{{\mu}{\nu}}=L_{{\mu} {\nu}} +S_{{\mu} {\nu}}. 
\label{4} 
\end{equation} 
is conserved. What can be known about particle
kinematics for a generic $S_{\mu \nu}$?.  $J$
is a constant, but whe have not proved the same for $L$ nor for $S$, we
only know that $\dot{S}=-\dot{L}$. The six constants of motion 
contained in $J$ can be made explicit by 
\begin{equation} X_{{\mu}}=\frac{1}{p^2}
J_{{\mu} {\nu}} p^{{\nu}},  \,  \, W_{{\mu}}=-J^*_{{\mu} {\nu}}
p^{{\nu}}=-S^*_{{\mu} {\nu}} p^{{\nu}} 
\label{5} 
\end{equation} 
 $W$ is the Pauli-Lubansky vector describing the spin
of the particle, while $X$ is a fixed position in terms of which 
we can express the trajectory as
\begin{equation}
x_{\mu}(\tau)=X_{\mu}+f(\tau)p_{\mu}+r_{\mu}(\tau)  
\label{11}        
\end{equation}   
The vector $r$, the radius, is defined by
\begin{equation}
r_{{\mu}} =- \frac{1}{p^2} S_{{\mu} {\nu}} p^{{\nu}}. 
\label{6} 
\end{equation} 
Its origin can be traced back to the independent existence of orbital 
and internal angular momentum. This vector yields a tiny deviation of 
order (spin/mass) from the trajectory of the spinless free particle
 Eq. (\ref{2}). It is not restricted to be a constant , opening the 
possibility for a non rectilinear $x(\tau)$ (a classical version of 
zitterbewegung). Sufficient coarsening will only reveal the ``external'' 
trajectory of (\ref{2}) and possibly a finite constant spin $W$.
Coarsening erases the internal angular momentum generated by 
zitterbewegung, the first term in 
\begin{equation}
 S_{{\mu} {\nu}}=-(r_{{\mu}}p_{{\nu}}-r_{{\nu}}p_{{\mu}})+
\frac{1}{p^2}\epsilon_{{\mu} {\nu} \rho \sigma} W^{\rho} p^{\sigma}.
\label{8} 
\end{equation} 

In general one would need of additional subsidiary conditions to fix the
time evolution of $r$. For instance, a constant spin modulus gives a constant
norm radius:
\begin{equation}
\frac{1}{2}S_{\mu \nu}S^{\mu \nu}=-\frac{W^2}{p^2}+p^2 r^2
\label{9}
\end{equation}
An arbitrary antisymmetric tensor decomposes into the sum of two orthogonal 
planes. The condition that $S_{\mu \nu}$ corresponds to one and only one 
of these planes is 
\begin{equation} 
\frac{1}{2}S^*_{{\mu} {\nu}}S^{{\mu} {\nu}}=2Wr=0 
\label{10} 
\end{equation}
Finally, r is a spacelike fourvector of constant norm, whose only allowed 
motion would be a rotation in the plane $\pi_r$ orthogonal to $W$ and $p$.
The particle would perform a helical motion composed of the uniform motion 
proper of the spinless particle and the rotation in the plane $\pi_r$.
The helix thread number is not given yet. A particular case of (\ref{9}) 
and (\ref{10}) is $S_{\mu \nu}\dot{x}^\nu=0$ which would add that information.

To describe the above situation beyond kinematics, one needs a lagrangian 
linear in velocities to avoid getting $\dot{x}$ in terms of $p$. This is
 not enough still, since the spin degrees of freedom have
to participate in the hamiltonian, and hence in the canonical formalism,
 in order to get a $r$ dependent trajectory. Here, we describe  classical
 spinning particles in mathematical terms as belonging to irreducible 
representations of the 
Poincare group of mass $m$ and spin $s$. We implement this idea in classical 
mechanics by including in the configuration space of the particle additional 
variables $\xi$
and $\dot{\xi}$ (functions of the proper time) transforming according to 
the representation chosen.
Invariance of the action under translations ($\delta \xi=0$) will be achieved
by momentum conservation, while Lorentz transformations ($\delta \xi=
\epsilon(s)\xi$) will require the conservation of total angular
momentum. The idea of using Dirac spinors in classical mechanics 
is originally due to Proca~\cite{proca} and 
Schiller~\cite{schi}. The latter  author
demonstrated that the hamiltonian proposed by Kramers~\cite{kramers} to 
describe the
evolution of a classical dipole in an external electromagnetic field could
be reformulated in terms of a Dirac bispinor. He then generalized the 
Hamilton-Jacobi formalism to include the motion of the dipole in a classical 
field theory framework. Proca intended to build a a new point mechanics in
terms of classical spinors to lay the foundations of a new quantization
programme. Barut and collaborators gave a vigorous impulse to this proposal 
-promoted by them to a classical model of the Dirac electron~\cite{barut1}- 
and were able 
to show the emergence of QED from the path integral of this classical
 particle~\cite{barut2}.

We now focus on the phase space of the spinning particle and recall that
 we are dealing with
elementary systems, i.e. with irreducible representations of the
Poincare Group~\cite{irreps}. The $\xi$'s will then be the spinors transforming
according to the representation chosen. To the pair of variables $\xi$
and its time derivative $\dot{\xi}$ we will associate a pair of
canonical conjugate variables $(\xi,\eta)$, and enlarge the
coordinates and momenta of the usual phase space to
$P =\{ ( x^\mu , p_\nu ),(\xi,\eta)\}$. Thus, each elementary
system is characterized by the representation chosen for the phase
space. $P$ can be labeled with two indices $(m,s)$ giving the
mass and spin of the particle. We will form Lorentz scalars, vectors 
and tensors out of these by
forming bilinears $\eta \Gamma_{\mu_1 \cdots \mu_n} \xi$ and combinations
of $x$ and $p$ in the standard form.We  also form higher spinors
from $\xi$ or $\eta$. We can give a simple recipe for the symplectic 
structure on
$P$: Given any pair of functions $A,B$ of $P$, we define
the canonical braket as
\begin{equation}
\{A,B\}=\frac{\partial A}{\partial x^{\mu}}\frac{\partial B}
        {\partial p_{\mu}}-
        \frac{\partial A}{\partial p_{\mu}}\frac{\partial B}
        {\partial x^{\mu}}+
        \frac{\partial A}{\partial \xi}\frac{\partial B}{\partial \eta}-
        \frac{\partial B}{\partial \xi}\frac{\partial A}{\partial \eta}
\label{12}
\end{equation}
where the order of the factors in the last term of the r.h.s. is
chosen by notational convenience. It can be seen easily that the above
fulfills all the conditions necessary to become an apropriate Poisson
bracket~\cite{dirac}.

We now require Poincare invariance: 
\begin{enumerate}
\item  Under translations
(the $p_{\mu}$ are constants, so the hamiltonian is independent of
$x$).
\item  Under Lorentz transformations (the $J_{\mu \nu}$ are constants,
so the hamiltonian must produce $\dot{S}=-\dot{L}$). 
\item Under parity
and time reversal (we will only consider representations $(j,j)$ or
$(j,0)\oplus (0,j)$ and their combinations for the Lorentz part, so $\eta \sim
\xi^{\dagger}\beta$ where $\beta$ is the parity matrix). 
\end{enumerate}
The solution to the equations
of motion have to be physically consistent. This can be translated
into three physical requirements: 
\begin{enumerate}
\item Physical trajectories are in
Minkowski space-time at all times. We  normalize the
proper time to the lenght of the trajectory, $\dot{x}^2=1$ . 
\item Physical variables  remain in the representation space chosen (the
invariants labeling the  representation have to be constants). 
\item Finally, there is only one object in phase space  (combined with
item 3. above this will lead to charge conjugation and CPT~\cite{wein1}).
\end{enumerate}
 We will not
analyze these issues in general, as they are well known~\cite{wein2} from
relativistic particle theory.

As said above, to disentangle momentum from velocity, one needs a
lagrangian linear in velocities. This singular case is best treated by the
method of Fadeev and Jackiw~\cite{jack} where the momenta are considered as
independent variables in configuration space, obtaining the same results than
the standard canonical treatment with lesser effort. We will use a 
generalization of the Proca lagrangian~\cite{proca} for arbitrary spin
\begin{equation}
L=i\frac{\lambda}{2}(\bar{\xi}\dot{\xi}-\dot{\bar{\xi}}\xi)
+\frac{1}{2}(p_{\mu}\dot{x}^{\mu}-\dot{p_{\mu}}x^{\mu})-
\bar{\xi}\beta_{\mu}p^{\mu} \xi \label{13}
\end{equation}
where $\beta$ is the irreducible part of
\begin{equation}
\beta_\mu =\frac{1}{2s}\left(\gamma_{\mu}\otimes I\otimes\ldots\otimes I+ 
\stackrel{2s}{\ldots}+I\otimes\ldots\otimes I\otimes\gamma_{\mu}\right)
\end{equation}
and $s$ is the spin label. The Euler Lagrange equations are: 
\begin{equation} 
\dot{x}^{\mu}=\bar{\xi}\beta^{\mu}\xi, \
\dot{p}_{\mu}=0,\, \ i\lambda \dot{\xi}=\beta\cdot p \xi, \  i\lambda
\dot{\bar{\xi}}=- \bar{\xi}\beta\cdot p \label{14} 
\end{equation} 
In (\ref{13},\ref{14}) $\lambda$ is an arbitrary scalar constant with the
dimensions of action. We have also taken the spinor momentum $\eta=i
\lambda \bar{\xi}$ without loss of generality. Observe how momentum 
and velocity are decoupled, 
the latter being fixed by the spinor degrees of freedom alone.

The spin tensor, Pauli-Lubansky vector, and radius are 
\begin{equation}
S^{\mu \nu}=-\lambda s \bar{\xi}\beta^{\mu \nu}\xi, \, 
r^{\mu}=
\frac{\lambda s}{p^2}\bar{\xi}\beta^{\mu \nu}\xi p_{\nu}, \,
W^{\mu}=\lambda s\bar{\xi} \beta^{* \mu \nu}\xi p_{\nu}
\label{15} 
\end{equation} 
where $\beta^{\mu \nu}= i s [\beta^\mu , \beta^\nu ]  $.
After some algebra one arrives to the Poisson brackets
\begin{eqnarray} 
\{W_{\mu},W_{\nu}\}&=&\epsilon_{\mu \nu \rho \lambda} p^{\rho}
W^{\lambda}, \ \{W_{\mu},r_{\nu}\}=\epsilon_{\mu \nu \rho \lambda}
p^{\rho} r^{\lambda}, \nonumber \\ \{r_{\mu},r_{\nu}\}&=&-
\frac{1}{p^4}\epsilon_{\mu \nu \rho \lambda} p^{\rho} W^{\lambda}
\label{16} 
\end{eqnarray} 
which show the close relation between these variables
and the rotations and boosts of the Lorentz group. We can use the
constant $W^{\mu}$ as giving the spin in the rest frame. Also, it is
possible to give the spin tensor (\ref{8}) in terms of $p$ and $r$
solely $S^{\mu \nu}=p^\mu r^\nu -p^\nu r^\mu-i p^2 \{r^\mu,r^\nu\}$.
Observe that $r$ can be made to vanish and still the Poisson Bracket
$\{r,r\}$ of (\ref{16}) be finite. 
Therefore, spin is not a consequence of zitterbewegung. We may have no 
zitterbewegung $(r=0)$, but finite spin ($W\neq 0$).
There is also a loose
relation between the product
of bilinears and the symplectic structure; here we are freed of the
problems~\cite{arlen1} that other approaches have with the simultaneous
 treatment~\cite{arlen2} of
classical and -the would be-  quantum variables.

The solution of the equations of motion (\ref{14}) will of course
lead to a trajectory of the form (\ref{11}). We get it explicitly
with 
\begin{equation}
X^\mu =x^\mu (0)-r^\mu (0),\
f(\tau)=\frac{1}{p^2}\bar{\xi}\beta\cdot p \xi \tau ,
\end{equation}
\begin{equation}
r^\mu (\tau)=r^\mu(0)\cos\left(\frac{\sqrt{p^2}\tau}{\lambda s}\right)+
\frac{\lambda s}{\sqrt{p^2}}\dot{r}^\mu (0)\sin\left(\frac{\sqrt{p^2}\tau}
{\lambda s}\right) \label{17}
\end{equation}
where $\dot{r}^\mu(0)=d^{\mu \nu}(p) \bar{\xi}\beta_{\nu}\xi$ and $r^\mu(0)$
is given by (\ref{15}). 
All the spinors in (\ref{17}) are given by their initial values at $\tau=0$ and
$d^{\mu \nu}(p)$ is the projector orthogonal to $p$. Equations
(\ref{2},\ref{17}) reveal the presence of zitterbewegung through a
radius joining the instantaneous centre of the particle ($X+fp$)
to its true position $x$. It is
the centre which follows the external trajectory along $p$ of a
spinless particle. The radius rapidly rotates around the centre (with the
frequency $2\sqrt{p^2}c^2/\hbar=1.5 \ 10^{21} s^{-1}$ for an electron and 
$\lambda=1$) tracing an ellipse whose parameters depend solely of the 
spinorial degrees of freedom of the particle. It is also worth to
note here that the particle does not spend any energy nor momentum 
in keeping zitterbewegung on. One
would also expect that classical external
perturbations would act adiabatically on this internal rotation. The
momentum  would change according to the applied external force as 
dictated by the Newton second law, but the particle would continue to
wind at each instant around the momentum to form a helix embracing the
external trajectory.

We will now show that the above picture is not consistent for charged
particles with spin. The reason is simply that zitterbewegung produces
a radiation field that should carry fourmomentum away from the free particle!. 
We first introduce the
electromagnetic interaction by the minimal substitution rule
$p\rightarrow p-eA$ in (\ref{13}). Then we obtain the electromagnetic
current $j^{\mu}=e \bar{\xi}\beta^{\mu}\xi=e u^{\mu}$. We
now recall that we can use all the results of the electrodynamics of
spinless particles that do not use of relations of the type $p^{\mu}=m
u^{\mu}$. Our spinning free particle will radiate at a rate
\begin{equation}
\frac{dp^{\mu}_{rad}}{d\tau}=-\frac{2}{3}e^2
(\ddot{r}_{\mu}\ddot{r}^{\mu})u^\mu
 \label{18} 
\end{equation}
Some comments are in order here: First,  radiation
will not drain four momentum off the particle ($\dot{p}=0$). Second,
radiation will occur even in
the ``rest system'' ($\vec{p}=0$) of the charge. Third, radiation will
not slow zitterbewegung, that proceeds at fixed frequency. The cure to these 
accumulation of catastrophes is to forbid
the free particle to radiate at all. Being $\ddot{r}$ spacelike, this can be 
achieved only if $\ddot{r}=0$. Inspecting (\ref{17}) we see that in
this case $r=0$ also. This is the condition of parallelism that can
be attained  when the spinor is one of the  eigenvectors
of the operator $\beta \cdot p$. For instance, for $s=1/2$, $r=0$ implies  
$\gamma\cdot p \xi_\pm =\pm \sqrt{p^2}\xi_\pm$. It is consistent~\cite{argens}
 to normalize $\bar{\xi}_\pm \xi_\pm =\pm 1$ and that when $p^0 <0$,
$\xi_{-}$ represents an antiparticle of momentum $-p$. Therefore, 
the condition of no spontaneous emision leads to
free particles being either particles or antiparticles, never a
mixture of these. Second cuantization is a natural outcome of the
above requisites. In addition, one recovers the old form of the Galileo 
principle.

The switch on of the electromagnetic interaction begins to complicate 
this very simple behaviour of the free spinning charged particle.
 Zitterbewegung will start with its
proper invariable frequency, but with a small growing amplitude  as the 
external
field feeds into a finite $r$ (for instance, for $s=1/2$ this will proceed 
by creating a finite $\xi_{-}$ ($\xi_{+}$)
component out of the initial $\xi_{+}$ ($\xi_{-}$)). Conversely, when exiting
from the interaction region the particle, generally off the mass shell,
 will radiate till reaching the pure state. One may think 
that the description in terms of spinors is a too ellaborated one to deal
with the spinning particle, at least in the free case. There may be a strong 
temptation to disentangle the radius from the Minkowski space coordinates. This
could be achieved by performing a canonical transformation from $x$ to $z=x-r$
getting rid of the cumbersome vanishing radius. However it is easy to check that
this new coordinate is not canonical; and the other way around: any canonical
 and covariant coordinate without the radius will acquire an imaginary part, 
which in addition 
turns out to be spin(-or) dependent. The lesson is clear: even if the kinematics
of the spinning particle traces that of the spinless case, there are 
additional degrees of freedom present, and these have to be taken into account
to implement the canonical formalism. In this sense, the study of spinning 
particles in ordinary $(x,p)$ phase space is hopeless.  

The work of J. M. Mart\'{\i}n was supported by Fundaci\'on Calvo Rod\'es and a 
grant from the Comunidad Aut\'onoma de Madrid.


\begin{thebibliography}{99}
\bibitem{corben} H. Corben, {\sl Classical and Quantum Theories of Spinning 
Particles}, Holden Day, Inc. San Francisco, 1968.
\bibitem{proca} A. Proca, J. Phys. Rad. 15, 65 (1954).
\bibitem{schi} R. Schiller, Phys. Rev. 128, 1402 (1962).
\bibitem{kramers} H. A. Kramers, Verhandl. Zeeman Jubil. 403 (1935).
\bibitem{barut1} A. O. Barut and N. Zanghi, Phys. Rev. Lett. 52, 2009 (1984).
\bibitem{barut2} A. O. Barut and I. H. Duru, Phys. Rev. Lett. 53, 2355 (1984),
                 and  Phys. Rep. 172, 1 (1989).
\bibitem{irreps} V. Bargmann and E. P. Wigner, Proc. N. A. S. 34, 211 (1948).
\bibitem{dirac} P. A. M. Dirac, {\sl Lectures in Quantum Mechanics}, Yeshiva
                University, New York, 1964. 
\bibitem{wein1} S. Weinberg, Phys. Rev. 133, B1318 (1964).
\bibitem{wein2} S. Weinberg, {\sl Quantum Theory of Fields}, Cambridge 
University Press, Cambridge, 1995.
\bibitem{jack} L. Fadeev and  R. Jackiw, Phys. Rev. Lett, 60, 1692 (1988).
\bibitem{arlen1} A. Anderson,  preprint quant-ph/9511014.
\bibitem{arlen2} A. Anderson,  Phys. Rev. Lett. 74, 621 (1995).
\bibitem{argens} F. H. Gaioli and E. T. Garc\'{\i}a Alvarez, 
                  Am. J. Phys. 63, 177 (1995).
\end{thebibliography}
\end{document}